\documentstyle[preprint,aps,version2,epsf]{revtex}
\def\bt{\beta}
\def\be{\begin{equation}}
\def\ee{\end{equation}}

\def\half{\frac{1}{2}}

\def\citen{\cite}
\def\ds{\displaystyle}

\begin{document}

\pagestyle{empty}
\preprint{RU--95--86}
\medskip
\preprint{hep-ph/9602021}
\medskip
\preprint{November 1995}

\vspace{-0.1in}

\begin{title}
\centerline{Manifolds of Fixed Points and Duality in}
\centerline{Supersymmetric Gauge Theories}
\end{title}

\author{
Matthew J. Strassler\footnote{E-mail address:
strasslr@physics.rutgers.edu}
\thanks{Work supported in part by the Department of Energy, contract
DE--FG05--90ER40559.}
}

\begin{instit}
Department of Physics and Astronomy\\
Rutgers University, Piscataway, New Jersey 08855
\end{instit}

\begin{abstract}

\vspace{-0.5in}

There are many physically interesting superconformal gauge theories
in four dimensions.   In this talk I discuss a common phenomenon in
these theories: the existence of continuous families of 
infrared fixed points.  Well-known examples include finite ${\cal N}=4$
and ${\cal N}=2$ supersymmetric theories; many finite ${\cal N}=1$
examples are known also.  These theories are a subset of a much
larger class, whose existence can easily be established and
understood using the algebraic methods explained here.  A relation 
between the ${\cal N}=1$ duality of Seiberg and duality in 
finite ${\cal N}=2$ theories is found using this approach, giving
further evidence for the former.  This talk is based on work
with Robert Leigh (hep-th/9503121).

\bigskip

\centerline{(To Appear in the Proceedings of the Yukawa
International Seminar [{\it YKIS} '95])}
\end{abstract}

\maketitle
\newpage

\pagestyle{plain}
\narrowtext

\section{Introduction}

  In this lecture, I will be presenting a new characterization of an 
important and common (but little known) phenomenon in four-dimensional
${\cal N}=1$ supersymmetric gauge theories.  This phenomenon, which goes 
by the name of ``manifolds of fixed points'' or 
``exactly marginal operators'', is familiar in two-dimensional field 
theory but has been little studied
in four dimensions.  A fixed point of a gauge theory is of course a
point in the space of coupling parameters where the theory is
conformally invariant; all coupling constants are truly constant, with
all beta functions vanishing and all of the physics scale-invariant.
A theory may have many isolated fixed points, but it may also possess
a continuous set --- a manifold --- of fixed points.  In this case,
starting from one such point, the Lagrangian of the theory can be 
smoothly deformed in such a way that the quantum theory remains scale
invariant.  The operator which deforms the Lagrangian is said to be 
an exactly marginal operator; in four dimensions, its dimension
is exactly four at every fixed point in the manifold.

  In general, dimensions of operators in quantum theories cannot
be computed exactly.  However, in certain theories, including
${\cal N}=1$ supersymmetric gauge theory in four dimensions, there is enough
symmetry in the theory that certain anomalous dimensions are exactly
known at fixed points, so operators can be shown to be exactly
marginal and the existence of manifolds of fixed points can be
confirmed.  This fact was proven in Ref.~\citen{rf:emop}, on which
this talk is based.

  The methods which are necessary for this purpose are remarkably
simple; they merely require algebra.  By this I do not mean abstract
group theory --- I mean the algebra we learned when we were children.
From these methods we can make exact statements about the behavior
of strongly coupled gauge theories, which allows us to clarify
many old results and leads to many new ones.  One application regards
the relation between the electric-magnetic duality of 
${\cal N}=1$ theories\cite{rf:sem} and that of ${\cal N}=2$ finite
theories\cite{rf:sw,rf:Ntwo}; it can be shown that the first follows
from the second.\cite{rf:emop}

  Some of our results on finite models are
related to earlier work of Lucchesi, Piguet and Sibold,\cite{rf:LPS}
though their methods are quite different from ours.

\section{Renormalization Group Flow}

Given a quantum field theory, which has gauge group $G$ and is weakly
coupled in the ultraviolet,  how does it change (flow)
under scaling transformations?  What
is its behavior in the infrared (IR)?  There are a number
of possibilities, which include:

\begin{enumerate}
\item No light particles in the IR.  The theory confines
or is completely Higgsed, and there is a mass gap.  

\item Only Goldstone bosons in the IR.  The theory confines
or is Higgsed, but global symmetries are broken spontaneously (as
in QCD).

\item Weakly coupled gauge theory, with gauge group $H\subset G$,
in the IR.  The theory is either partly broken (as in the
electroweak interactions) or partly confined (with light
composite scalars and fermions potentially present.)

\item Weakly coupled gauge theory, with some other gauge group $G'$,
in the IR.
The gauge group $G$ is confined and a new gauge group emerges,
with composite gauge bosons and matter emerging at low energies.

\item Interacting conformal fixed point in the IR.  Many gauge theories
may flow to the same such point.
\end{enumerate}
Answers 4 and 5 are the most exotic, and only recently has it been
realized\cite{rf:sem,rf:kinsrev} that these are in fact common 
situations in supersymmetric gauge theory.

\section{Theories with Fixed Points.}

I now turn to the specific issue of theories with IR fixed points.
The fact that there are gauge theories in four dimensions
that have non-trivial interacting fixed points has
long been known.  One need only consider the
beta function for QCD with $N_c$ colors and $N_f$ flavors
of quarks in the fundamental representation:
\be\label{eq:betaQCD}
\begin{array}{rcl}
\beta(g) =-\frac{\ds g^3}{\ds 16\pi^2}
{\ds \sum_0^\infty \left[\frac{g^2}{16\pi^2}\right]^{k}b_k\ };& \\
b_0 = \frac{11}{3} N_c- \frac{2}{3}N_f \ ;  
b_1 = \frac{N_c}{3}&\left[34N_c -\left(13-\frac{3}{N_c^2}\right)N_f\right]\ .
\end{array}\ee
Here $b_n$ is the $n$-loop contribution to the beta function.
For asymptotically free theories, all the coefficients $b_k$ are of 
order at most $(N_c)^{k+1}$. If we take $N_f$ very close to but less than 
$\frac{11}{2}N_c$, for $N_c$ large (though not infinite), then the
leading term in the beta function will be negative and of order $1$ while
the next-to-leading term will be positive and of order $g^2 N_c^2$.
The beta function therefore appears to have a zero near 
$g\sim {4\pi}/{N_c}$.  Can we trust this zero?  Yes; for
sufficiently large $N_c$ this value of the coupling is sufficiently
small that the higher-order terms in the expansion of $\bt(g)$
will not change its basic functional form at small $g$.  We
may therefore conclude that
the beta function is negative at very small $g^2$ but positive
beyond some value.  At the zero of the beta function lies
a weakly interacting conformal field theory.  

This same argument applies in supersymmetric QCD; for
$N_f$ just less than $3N_c$ there are fixed points at weak coupling.  In
fact, Seiberg has argued\cite{rf:sem} (and by now the evidence is 
overwhelming\cite{rf:emop,rf:kinsrev}) that such interacting
field theories exist even for $N_f$ much less than $3N_c$, with
couplings which in general are non-perturbatively strong.

Another theory which has conformal fixed points
is ${\cal N}=4$ supersymmetric gauge theory.  In fact, this
theory is conformal for every value of the
gauge coupling.  Let us consider a slightly more general theory;
take an ${\cal N}=1$ supersymmetric gauge theory, with three chiral 
superfields $\phi_1,\phi_2,\phi_3$ in the adjoint representation of the 
group, and with a superpotential
\begin{eqnarray}\label{eq:Nisfour}
W= h f_{abc} \phi_1^a\phi_2^b\phi_3^c
\end{eqnarray}
where we do not assume $h$ and the gauge coupling $g$ are equal.  
In the space of all possible couplings 
$(g,h)$ there are non-trivial beta functions $\beta_g(g,h)$ and
$\beta_h(g,h)$; but on the line $h=g$ both beta functions
vanish by ${\cal N}=4$ supersymmetry.  Thus, in this 
${\cal N}=1$ theory there is a line of fixed points, labelled by $g$;
it can easily be shown that these fixed points are infrared stable.  
(In fact, for $SU(N)$ there
is a much larger manifold of fixed points which contains ${\cal N}=4$ as a 
subspace; see Ref.~\citen{rf:emop}.)
The variation of the  Lagrangian with respect
to $g$, $\frac{d {\cal L}}{d g}$, is an exactly marginal operator
on the line $h=g$.

\firstfigfalse
\centerline{\leavevmode\epsfysize=3in  \epsfbox{figa.epsf}}
\figure{\label{fig:Nisfour1}
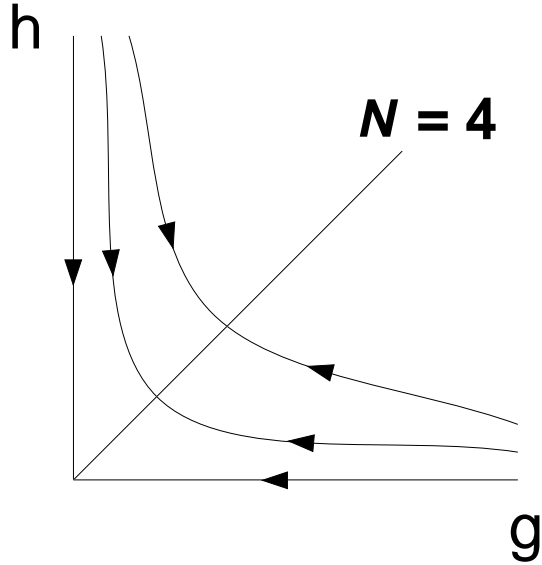
The ${\cal N}=4$ superconformal theories lie on a line 
inside the ${\cal N}=1$ theory (\ref{eq:Nisfour}). Arrows indicate
    renormalization group flow toward the infrared.}

The reason for viewing the ${\cal N}=4$ theory as a line inside the space
of couplings of an ${\cal N}=1$ theory is the following: there are many
examples, known since the 1980s, in which an ${\cal N}=1$ theory has
a curve of fixed points inside the space of couplings, even though
there is no special symmetry on this curve.  From this point of
view, ${\cal N}=4$ is a special case of a much broader class of ${\cal N}=1$ 
theories which have neither beta functions nor anomalous dimensions for their
chiral superfields; their effective actions are 
finite.\cite{rf:finiteearly,rf:listfin,rf:finitelate}  These
theories were the motivation for the Principle of Reduction of
Couplings\cite{rf:RoC} mentioned in the talk of Dr.~Kubo.  

I will now show that there is a simple way to understand
why these theories exist, and that there are many theories
which are {\it not} finite which nonetheless have manifolds
of fixed points.

\section{Proving an Operator is Marginal}

The question I will now address is this: given a theory which
is at a fixed point, whose Lagrangian is ${\cal L}$, 
and which has an operator ${\cal O}$ which is marginal (dimension-four)
at the fixed point, under what circumstances is the theory
with Lagrangian ${\cal L}+\epsilon{\cal O}$ still at a fixed point?
This is equivalent to asking whether the dimension of operator 
${\cal O}$ is still four when $\epsilon$ is non-zero.  If the
answer to this question is yes, then the operator is exactly marginal.

In general, this is a very difficult question to answer, even in
two dimensions, unless the theory is exactly soluble.  But
remarkably, it is easy to answer this question for certain operators
in ${\cal N}=1$ $d=4$ supersymmetry. This follows from three important and
little known facts about these theories:

\begin{enumerate}
\item Anomalous dimensions $\gamma$ of charged matter fields
are gauge invariant.

\item If a term $h\phi_1\dots\phi_k$ appears in the superpotential,
then
\be\label{betah} 
\beta_h \propto \left[(k-3)+\half\sum_{i=1}^k \gamma_i\right]
\ee
where $\gamma_i$ is the anomalous mass dimension of the field $\phi_i$.

\item  For the gauge coupling $g$, the beta function, in terms
of the second Casimir invariant of the gauge group $C_2(G)$
and the index $l_R=2T(R)$ of the representation $R$, is
\be\label{betag}
\bt_g \propto -\bigg(\Big[3C_2(G)- \sum_i T(R_i)\Big]
                   +\sum_i T(R_i) \gamma_i\bigg)
\ee
where the sums are over all matter fields.
Notice the one-loop beta function coefficient $b_0=3C_2(G)- \sum_i T(R_i)$ 
appears in this formula.
\end{enumerate}
The expression (\ref{betah}) follows directly from the non-perturbative 
non-renormalization theorem\cite{rf:Holo} for the superpotential, 
while (\ref{betag})
was proven to all orders in perturbation theory by Shifman
and Vainshtein\cite{rf:sv} (see also papers with Novikov and 
Zakharov\cite{rf:svnz}).  The latter is believed also to hold 
non-perturbatively
because the zeroes of the beta function are related, by properties of
the superconformal algebra of ${\cal N}=1$ 
supersymmetry,\cite{rf:Scurrents,rf:Sanomalies} to certain chiral
anomalies, which can be computed at one loop.

What is critical to note about these beta functions is that both
are linear functionals of the anomalous dimensions $\gamma_i$
of the matter fields $\phi_i$.  These anomalous
dimensions are arbitrarily complicated functions of the
coupling constants, but the  relations between beta functions
and anomalous dimensions are very simple.  We will now see
how this leads to a powerful and elegant result.

The criterion for the existence of a fixed point is that
all beta functions must vanish simultaneously.  The equations
$\beta_i=0$ put $n$ conditions on $n$ couplings.  Generically,
we would expect these conditions are satisfied at isolated points
in the space of couplings.  However, because the beta functions
are linear in the anomalous dimensions, it may happen that some
of them are linearly dependent.  If only $p$ of the beta functions
are linearly independent, then the equations $\beta_i=0$ put
only $p$ conditions on the $n$ couplings, so we expect that 
solutions to the equations generically occur on $n-p$ dimensional 
subspaces of the space of couplings, leading to $n-p$ dimensional 
manifolds of fixed points.

Thus, if some of the beta functions are linearly dependent,
and there is at least one generic fixed point somewhere in
the space of couplings, then the theory will have a manifold of
fixed points. Of course, it is possible that a theory has no 
fixed points at all (other than the free fixed point, which is not 
generic.)  But in many theories, non-trivial fixed points are known 
or believed to exist, and in some cases these extend to entire manifolds.

\section{Examples}

As a first example of such a theory, consider $SU(3)$ with
fields $Q^r,\tilde Q_s$ $(r,s=1,\dots,9)$ in the ${\bf 3}$
and ${\bf\overline 3}$ representations, with a superpotential
\be\label{suthreepot}
W=h\left(
Q^{1}Q^{2}Q^{3}+Q^{4}Q^{5}Q^{6}+Q^{7}Q^{8}Q^{9}+
\tilde Q_{1}\tilde Q_{2}\tilde Q_{3}+
\tilde Q_{4}\tilde Q_{5}\tilde Q_{6}+
\tilde Q_{7}\tilde Q_{8}\tilde Q_{9}\right)
\ .
\ee
The special properties of this theory were first noted in the
1980s.\cite{rf:finiteearly,rf:listfin}  Notice that the one-loop 
gauge beta function is zero and that the superpotential preserves
enough of the flavor symmetry that all $Q^r$ and $\tilde Q_s$
have the same anomalous dimension $\gamma(g,h)$.
The beta functions for the Yukawa coupling and gauge coupling are
\be\label{suthreebta}
\bt_h\propto \frac{3}{2}h\gamma(g,h); \ 
\bt_g\propto -9\gamma(g,h) \ .
\ee
For both of these to vanish requires only the one condition 
$\gamma(g,h)=0$.  Since $\gamma(0,0)=0$, $\gamma(g,0)<0$
and $\gamma(0,h)>0$ for small $g,h$, we conclude there is
a line of fixed points passing through the origin of the space
of couplings $(g,h)$.
  
\centerline{\leavevmode\epsfysize=3in  \epsfbox{figb.epsf}}
\figure{\label{fig:suthree}
 The renormalization group flow near the fixed curve associated
with the superpotential (\ref{suthreepot}), shown
schematically, in the plane of the gauge coupling $g$ and the Yukawa
coupling $h$.}
\noindent There are a number of important facts worthy of note:
\begin{enumerate}

\item On the fixed line, by definition, the theory is superconformal.
Since all beta functions and anomalous dimensions vanish, the
effective action on the line is finite (just as ${\cal N}=4$ theories are
finite.)

\item We have not determined the function $\gamma(g,h)$, but
we didn't need to.  The existence and properties of
the fixed line only required that $\gamma(g,h)$ be zero 
somewhere.  Furthermore, if $g,h$ are small,
we can compute $\gamma(g,h)$ in perturbation theory, 
while if $g,h$ are large this function is not useful anyway.

\item The line of fixed points is IR stable; if the 
couplings are near but off the line, they will approach the
line at low energy.  Therefore, no fine tuning is required to put
the theory on the line, which means these superconformal theories are
physically interesting.\cite{rf:RoC}

\end{enumerate}
There are an enormous number of  
models with similar manifolds of fixed points passing through zero
coupling; these can be and were identified in perturbation 
theory.\cite{rf:finiteearly,rf:listfin,rf:finitelate}

As another example, consider an ${\cal N}=1$ supersymmetric
$SU(N)$ gauge theory with fields $Q^r,\tilde Q_s$ $(r,s=1,\dots,N_f)$
in the fundamental and antifundamental representations, a single field
$\Phi$ in the adjoint representation, and a superpotential
\be\label{NtwoW}
W=h\Phi Q^r\tilde Q_r
\ee
where the repeated index is summed over.
For $h=g$ this theory has ${\cal N}=2$ extended supersymmetry.
In the case $N_f=2N$, the theory is finite on the line $h=g$; we 
can see this line of fixed points must exist using the fact that
both $\bt_g$ and $\bt_h$ are proportional to $\gamma_\Phi+2\gamma_Q$
(by symmetry, all $Q^r,\tilde Q_s$ have the same anomalous dimension).
Again the line of fixed points is infrared stable.

But now let us add a mass for the field $\Phi$ and integrate it
out.  The low energy theory is an ${\cal N}=1$ $SU(N)$ gauge theory with the
$N_f=2N$ fields $Q^r,\tilde Q_s$ coupled by a non-renormalizable 
superpotential
\be\label{NtwoslW}
W=\half m\Phi^2 + h \Phi Q^r\tilde Q_r \rightarrow 
W_L = \half y\left[(Q^r\tilde Q_s)(Q^s\tilde Q_r)
       -\frac{1}{N}(Q^r\tilde Q_r)(Q^s\tilde Q_s)\right]\ ,
\ee
where an $SU(N)$ Fierz identity has been used to write the low
energy superpotential $W_L$ in terms of gauge-invariant bilinears.
This superpotential preserves enough flavor symmetry that all
$Q^r,\tilde Q_s$ have the same anomalous dimension $\gamma(g,y)$.
The beta function for $y$ is $\bt_y\propto 1+2\gamma$ while
that of the gauge coupling is $\bt_g \propto (3N-N_f)+N_f\gamma$,
which for $N_f=2N$ is proportional to $\bt_y$.  We therefore
have a fixed point whenever $\gamma(g,y)=-\half$.

When the couplings $g,y$ are small, $\gamma\sim Ay^2-Bg^2$ for positive
coefficients $A,B$; $|\gamma|\ll\half$ at weak coupling, and any
fixed point is therefore non-perturbative.  The work of 
Seiberg\cite{rf:sem} strongly suggests that there is a value of $g=g_*$
for which the theory with $W_L=0$ is at a stable
fixed point.  If this is true, then we expect
that $\gamma(g,0)>-\half$ [$\gamma(g,0)<-\half$] for $g<g_*$ [$g>g_*$].
There must therefore be a line of stable fixed points, passing through
the Seiberg fixed point $(g,y)=(g_*,0)$, which all have 
$\gamma(g,y)=-\half$. 

\centerline{\leavevmode\epsfysize=3in  \epsfbox{figc.epsf}}
\figure{\label{fig:Ntwosl}
The renormalization group flow near the fixed curve associated
with the superpotential (\ref{NtwoslW}), shown
schematically, in the plane of the gauge coupling $g$ and the Yukawa
coupling $h$.}

\begin{enumerate}

\item The fact that this theory has a quartic, non-renormalizable
superpotential poses no difficulty.  The low-energy theory is
an effective theory valid below the mass of $\Phi$.  Above this
mass the theory is renormalizable.   In fact, one may redo the
analysis in the renormalizable theory and obtain the same result
concerning the existence of a fixed line.

\item  The coupling $y$ in (\ref{NtwoslW}) has dimension of inverse mass,
so one might ask how it can appear in a superconformal field theory.
The reason is that the anomalous dimension of the operator $(Q\tilde Q)^2$
is $\half\cdot 4\gamma=-1$ at the fixed point; this exactly cancels the 
naive dimension of the coupling $y$.

\item As before, the fixed line is IR stable, so no fine-tuning
is required for the theory to be superconformal in the IR. 

\end{enumerate}
Again, the number of similar examples is enormous.\cite{rf:emop}

\section{Application to ${\cal N}=1$ Duality}

While there is insufficient space to present the full relation
between ${\cal N}=2$ and ${\cal N}=1$ duality, I will sketch the main 
ideas;\footnote{This section was
omitted from the talk due to time constraints.}
further discussion is to be found in Ref.~\citen{rf:emop}.
(In intervening months, an improved understanding of this
relation has been achieved; more up-to-date discussion can
be found in Ref.~\citen{rf:NtwoNone}.)

The ${\cal N}=2$ finite $SU(N)$ theory, whose line of fixed points 
was discussed above, has the property that at every
point along the ${\cal N}=2$ fixed line, there are (at least) two 
equivalent descriptions of the superconformal theory.\cite{rf:sw,rf:Ntwo}  
We may refer to the two sets 
of variables as electric and magnetic.   In both variables the model is a
 finite ${\cal N}=2$ $SU(N)$ theory, but the magnetic  coupling $\tilde g$
is the inverse of the electric coupling $g$, and the flavor quantum 
numbers are assigned differently in the two descriptions.

When a mass is given to the field $\Phi$ (this occurs similarly
in both sets of variables) both descriptions flow to a low-energy
theory consisting of an ${\cal N}=1$ $SU(N)$ gauge theory with $N_f=2N$,
as above.  Both theories have a superpotential of the form (\ref{NtwoslW}),
but the coupling $\tilde y$ in the magnetic description is the
inverse of $y$ in the electric description.  If we take the coupling $g$
of the ${\cal N}=2$ electric theory to be arbitrarily small, then the 
electric $y$ will be arbitrarily small as well; in the
limit, the ${\cal N}=1$ electric superpotential is zero, but the theory will
still flow from weak coupling $(g=0)$ to its fixed point $(g=g_*)$
since it is asymptotically free.   

On the other hand, for large $\tilde g$ the magnetic theory will be at
strong coupling, and will have a superpotential 
with a huge coefficient $\tilde y$.  But we may introduce an auxiliary gauge
singlet field $M^r_s$ to rewrite it as follows.
\be\label{MW}
W_L^{mag} = -\half \tilde y\left[(q^r\tilde q_s)(q^s\tilde q_r)
       -\frac{1}{N}(q^r\tilde q_r)(q^s\tilde q_s)\right]
\rightarrow
 M^r_s q_r\tilde q^s + 
\frac{\ds 1}{\ds 2\tilde y}(M^r_sM^s_r-\frac{1}{N}M^r_rM^s_s)
\ .
\ee
In the limit that the magnetic theory is infinitely coupled
(which is the same limit we considered above for the electric theory)
the mass of the meson $M$ is zero and we are left with the
superpotential
\be\label{magW}
W_L^{mag}=M^r_s q_r\tilde q^s \ .
\ee
We expect the low-energy ${\cal N}=1$ magnetic theory to flow from infinite
coupling back to its stable IR fixed point\cite{rf:sem} 
$\tilde g = \tilde g_*$.

We therefore conclude that the duality of the ${\cal N}=2$ theory implies that
an ${\cal N}=1$ $SU(N)$ gauge theory with $N_f=2N$ and superpotential $W=0$
is dual to another $SU(N)$ theory with $N_f=2N$ and mesons $M^r_s$
coupled by the superpotential $W=Mq\tilde q$.  This is exactly
what was found by Seiberg in Ref.~\citen{rf:sem}.

More details of this relation may be checked.\cite{rf:emop}  In
particular, the mapping between operators found in Ref.~\citen{rf:sem}
is also recovered.  Using the mapping, one can flow away from the
theory with $N_f=2N$ and show that the $SU(N)$ theory with $N_f$
flavors has a magnetic dual which has $SU(N_f-N)$ gauge group,
$N_f$ flavors, singlets $M^r_s$, and the 
superpotential (\ref{magW}).\cite{rf:emop}

\section{Conclusions}

\begin{enumerate}

\item Manifolds of fixed points are a 
 common phenomenon in ${\cal N}=1$ supersymmetric gauge theories.

\item These manifolds can easily be found, using simple algebraic methods.

\item The ``finite'' ${\cal N}=1$ models discovered ten years ago are
easily understood using this language.  Proving a model is finite is 
often trivial with these methods.

\item Many ``non-finite'', non-renormalizable, strongly coupled
manifolds of fixed points have also been uncovered using this approach.

\item These fixed points are usually IR stable, and are
therefore of physical interest.

\item Manifolds of fixed points appear to play an important
role in duality, as manifested by the relation discussed above
between duality in finite ${\cal N}=2$ models and
Seiberg's ${\cal N}=1$ duality.

\item Beyond those involving the Principle of Reduction of 
Couplings\cite{rf:RoC}, no applications to phenomenology have yet been 
suggested, but approximate fixed points have the potential to play
an important role in supersymmetric model building. 
\end{enumerate}

\section{Acknowledgements}{I am grateful to the organizers 
for having given me the opportunity to attend such an 
interesting and exciting conference.}

\newcommand\bib[5]{#1 \it #2\bf#3\ \rm(19#5)\ #4}
\newcommand\bibh[7]{#1 \it #2\bf#3\ \rm(19#5)\ #4, hep-#6h/#7}
\newcommand{\PRL}{Phys. Rev. Lett. }
\newcommand{\PLB}{Phys. Lett. \bf B}
\newcommand{\PRD}{Phys. Rev. \bf D}
\newcommand{\NPB}{Nucl. Phys. \bf B}
\newcommand{\IJMPA}{Int. J. Mod. Phys. \bf A}

\end{document}